\documentclass[twoside,english,aps, tightenlines,showpacs, showkeys]{revtex4-2}
\usepackage[latin9]{inputenc}
\usepackage{color}
\usepackage{babel}
\usepackage{amsmath}
\usepackage{esint}
\usepackage[unicode=true,pdfusetitle,
 bookmarks=true,bookmarksnumbered=false,bookmarksopen=false,
 breaklinks=false,pdfborder={0 0 0},pdfborderstyle={},backref=false,colorlinks=true]
 {hyperref}

\makeatletter

\newcommand{\lyxmathsym}[1]{\ifmmode\begingroup\def\b@ld{bold}
  \text{\ifx\math@version\b@ld\bfseries\fi#1}\endgroup\else#1\fi}

\DeclareTextSymbolDefault{\textquotedbl}{T1}

\usepackage{dcolumn}
\usepackage{bm}

\makeatother

\begin{document}
\title{Maxwell extension of $f(R)$ gravity }
\author{Oktay Cebecio\u{g}lu$^{1}$}
\email{ocebecioglu@kocaeli.edu.tr}

\author{Ahmet Saban$^{1}$}
\email{ahmetsaban55@gmail.com}

\author{Salih Kibaro\u{g}lu$^{2,3}$}
\email{salihkibaroglu@maltepe.edu.tr}

\date{\today}
\begin{abstract}
Inspired by the Maxwell symmetry generalization of general relativity
(Maxwell gravity), we have constructed the Maxwell extension of $f(R)$
gravity. We found that the semi-simple extension of the Poincare symmetry
allows us to introduce geometrically a cosmological constant term
in four-dimensional $f(R)$ gravity. This symmetry also allows the
introduction of a non-vanishing torsion to the Maxwell $f(R)$ theory.
It is found that the antisymmetric gauge field $B^{ab}$ associated
with Maxwell extension is considered as a source of the torsion. It
is also found that the gravitational equation of motion acquires a
new term in the form of an energy-momentum tensor for the background
field. The importance of these new equations is briefly discussed.
\end{abstract}
\affiliation{$^{1}$Department of Physics, Kocaeli University, 41380 Kocaeli, Turkey,}
\affiliation{$^{2}$Department of Basic Sciences, Faculty of Engineering and Natural
Sciences, Maltepe University, 34857, Istanbul, Turkey}
\affiliation{$^{3}$Institute of Space Sciences (CSIC-IEEC) C. Can Magrans s/n,
08193 Cerdanyola (Barcelona) Spain}
\maketitle

\section{Introduction}

Although general relativity (GR) is widely accepted as a fundamental
theory to describe the gravitational phenomena on an astrophysical
scale, it does not explain for the rotational curves of galaxies that
have been measured do not fit the predictions of GR with baryonic
matter and predict the accelerated expansion of the universe that
was observed at the end of the last century \citep{riess1998observational}.
The explanation in the case of rotational curves is to introduce a
new type of non-baryonic matter (dark matter) \citep{Trimble:1987Existence,Padmanabhan:2003Cosmological}.
The accelerated expansion of the universe is usually explained by
invoking a mysterious substance called dark energy. The simplest candidate
for dark energy is the cosmological constant \citep{Peebles:2003Cosmological}.
Needless to say, the cosmological constant problem is one of the major
challenges in theoretical physics. Introducing mysterious substances
to match experimentally observed values with the theoretical predictions
of GR is one of the approaches to overcome the problem. In this approach,
one modifies the matter part of the Einstein field equations. Another
approach is to modify the left-hand side (geometric part) of the Einstein
field equation, called as a modified gravitational theory, in which
the standard Einstein-Hilbert (E-H) action is replaced by an arbitrary
function of the Ricci scalar $R$. Such a modification first was put
forward by Buchdahl in 1970 \citep{Buchdahl:1970Non}. This theory
is called today $f(R)$ gravity and became an established field of
theoretical gravity and cosmology after the influential work by Starobinsky
\citep{Starobinsky:1980ANew}. The current acceleration of the universe
can be explained by $f(R)$ gravity \citep{nojiri2003modified,Capozziello:2003quintessence,Carroll:2004IsCosmicSpeed,nojiri2007introduction,Hu:2007ModelsOfF(R),Appleby:2007DoConsistent,Starobinsky:2007DisappearingCCF(R)}.
Viable models of dark energy satisfying the Solar system and cosmological
observational data based on $f(R)$ gravity where $f(R)$ is finite
at $R=0$ were first independently constructed in \citep{Hu:2007ModelsOfF(R),Appleby:2007DoConsistent,Starobinsky:2007DisappearingCCF(R)}
and previous models where $f(R)$ diverges at $R=0$ were shown to
be not viable in \citep{Amendola:2007Conditions}. For more information
as well as recent developments and their applications to the physically
relevant models of $f(R)$ theories, see one of the excellent reviews
\citep{Sotiriou:2007MetricAffine,Sotiriou:2010FR,nojiri2011unified,DeFelice:2010FR,Capozziello:2011Extended,Clifton:2011Modified,Olmo:2011uz,Nojiri:2017Modified,Heisenberg:2019systematic}
and references therein.

There exists another interesting class of modified gravity theory
which may easily produce the cosmological constant by gauging the
Maxwell algebra, so-called Maxwell-gravity \citep{azcarraga2011generalized}.
Starting with the work of Bacry et al. \citep{bacry1970group,bacry1970group2},
the idea of Maxwell symmetry has been systematically studied by Schrader
\citep{schrader1972maxwell}. Such a symmetry describes a charged
particle moving in a four-dimensional Minkowski background in the
presence of a constant electromagnetic field. The Maxwell algebra
is an extension of the Poincare algebra by six additional tensorial
abelian symmetry generators that make the four-momenta non-commutative
$[P_{a},P_{b}]=i\lambda Z_{ab}$ \citep{soroka2005tensor}. In 2012,
the semi-simple tensor extension of the Poincare group was given by
Soroka with a new non-abelian tensorial generator \citep{soroka2012gauge}.
In this study, another alternative approach to the cosmological term
problem is proposed. After the work of Azcarraga and Soroka, there
has been a renewed interest in the cosmological constant problem due
to Maxwell symmetry. Various studies on the gauge theory of the (super)
Maxwell symmetry algebras carried out and different aspects has been
studied in \citep{Gomis:2009Deformations,bonanos2009infinite,Bonanos:2009ChevalleyEilenberg,concha2014maxwell,cebeciouglu2014gauge,cebeciouglu2015maxwell,concha2015generalized,kibarouglu2019maxwellSpecial,kibarouglu2019super,kibarouglu2020generalizedConformal,kibarouglu2021gaugeAdS,cebeciouglu2021maxwellMetricAffine}. 

As is well known, the simplest candidate for describing dark energy
is the cosmological constant. Then, it becomes interesting to study
the ways in which cosmological constant terms can be introduced in
the $f(R)$ theories. In particular, the $f(R)$ theory of gravity
with a cosmological term can be developed in a geometric formulation,
where the theory is constructed from the curvatures of the semi-simple
extended Poincare algebra. The main purpose of the present paper is
to generalize the metric $f(R)$ theory to a situation with extra
degrees coming from Maxwell symmetry extension. In this way, we give
an alternative way of introducing the cosmological term to the $f(R)$
theory. 

The paper is organized as follows: In Section II, we briefly recall
the construction of $f(R)$ theory and summarize its basic equations
in both metric and Palatini formalism. In Section III, we present
the construction of a four-dimensional gravity model containing a
cosmological constant term only from the (linear in) curvatures of
semi-simple Poincare algebra, the action we reached corresponds to
an Einstein-Hilbert like action. In Section IV, we propose our simple
model of the Maxwell generalized $f(R)$ gravity action involving
a cosmological term and comment on the obtaining of the equations
of motion. Finally, in Section V, we discuss the obtained results
and possible future developments.

\section{$f\left(R\right)$ theory in terms of differential forms}

There are three versions of $f(R)$ gravity: Metric-$f(R)$ theory
(second order formalism) is fully described by the metric field alone,
Palatini-$f(R)$ theory (first order formalism) in which metric and
connection are handled as independent fields, similarly the \textquotedbl metric-affine
$f(R)$ theory\textquotedbl{} in which matter Lagrangian also includes
connection \citep{Sotiriou:2007MetricAffine}. In this section, we
very briefly summarize the main ingredients of $f(R)$ gravity in
both metric and Palatini approaches. The starting point for the metric-$f(R)$
gravity is the Einstein-Hilbert action, 

\begin{equation}
S_{EH}=\frac{1}{2\kappa}\intop d^{4}x\sqrt{-g}R,\label{eq:EH-action}
\end{equation}
in which $\kappa=8\pi G/c^{4}$ with $G$ being Newton's gravitational
constant and the Ricci scalar $R$ is constructed from the Riemann
curvature tensor. One of the simplest modifications to general relativity
is the $f(R)$ gravity. It generalizes the Lagrangian density of the
E-H action. Specifically, it replaces the Ricci scalar $R$ in action
Eq.\eqref{eq:EH-action}, with some function $f(R)$ of the scalar
curvature:

\begin{equation}
S_{f(R)}=\frac{1}{2\kappa}\intop d^{4}x\sqrt{-g}f(R).\label{eq:f(R) action}
\end{equation}
The source-free vacuum field equations that are obtained taking the
variations of action with respect to the metric $g_{\mu\nu}$ are

\begin{equation}
f^{\,'}R_{\mu\nu}-\frac{1}{2}fg_{\mu\nu}+(g_{\mu\nu}\Delta-\nabla_{\mu}\nabla_{\nu})f^{\,'}=0,\label{f(R)-tensor}
\end{equation}
where $\Delta=g{}^{\mu\nu}\nabla_{\mu}\nabla_{\nu}$ is the d\textquoteright Alembertian
operator. Note that the cosmological constant term does not appear
in this field equation. These equations can be re-arranged in the
Einstein-like form

\begin{equation}
G_{\mu\nu}=R_{\mu\nu}-\frac{1}{2}g_{\mu\nu}R=T_{\mu\nu}^{eff},
\end{equation}
where

\begin{equation}
T_{\mu\nu}^{eff}=\frac{1}{f^{\,'}}\left\{ \nabla_{\mu}\nabla_{\nu}f^{\,'}-g_{\mu\nu}\Delta f^{\,'}+\frac{1}{2}g_{\mu\nu}\left(f-f^{\,'}R\right)\right\} ,
\end{equation}
is an effective energy-momentum tensor which can be interpreted as
an extra gravitational energy-momentum tensor due to higher order
curvature effects. Including the function $f(R)$ gives extra freedom
in defining the behavior of gravity. The detailed structure of $f(R)$
gravity theories arising from the action Eq.\eqref{eq:f(R) action},
in $4D$ space-time has been discussed in ref. \citep{Sotiriou:2007MetricAffine,Sotiriou:2010FR,DeFelice:2010FR,Capozziello:2011Extended,Clifton:2011Modified,Nojiri:2017Modified}.

In the context of the first order (Palatini) formalism, the entities
basis 1-forms $e^{a}$ and the connection 1-forms $\omega^{ab}$ are
independent from each other. That is, the connection is not previously
fixed to be given by Christoffel\textquoteright s symbols, but must
be determined dynamically. Consequently, the general form of the field
equations does not forbid the presence of torsion \citep{Clifton:2011Modified}.
In the language of exterior differential forms, E-H action takes the
following form

\begin{equation}
S_{EH}=\frac{1}{2\kappa}\intop\mathcal{L}_{EH},
\end{equation}
with the gravitational Lagrangian 4-form is given by 

\begin{equation}
\mathcal{L}_{EH}=\frac{1}{2}\varepsilon_{abcd}R^{ab}\wedge e^{c}\wedge e^{d}=R^{ab}\wedge{}^{*}e_{ab}=R{}^{*}1,\label{EH-lag}
\end{equation}
where we denote the exterior products of basis 1-forms $e^{a}$ as
$e^{ab}=e^{a}\wedge e^{b}$, $e^{abc}=e^{a}\wedge e^{b}\wedge e^{c}$
etc., $R^{ab}$ is the Ricci 2-form, $^{*}$ denotes the Hodge dual
operator acting on the basis forms and $^{*}1=\frac{1}{4!}\varepsilon_{abcd}e^{abcd}$
is the invariant volume element. 

We consider the modified action 

\begin{equation}
S_{f(R)}=\frac{1}{2\kappa}\intop\mathcal{L}_{f(R)},
\end{equation}
and take the following Lagrangian 4-form in differential form language:

\begin{equation}
\mathcal{L}_{f(R)}=f(R)^{*}1.
\end{equation}
The full variation of the 4-form $\mathcal{L}_{f(R)}$ can be calculated
as follows. Using the product rule for the variation gives 

\begin{equation}
\mathcal{\delta L}_{f(R)}=\delta f(R){}^{*}1+f(R)\left(\delta{}^{*}1\right)=f^{\,'}\delta(R^{*}1)+\left(f-f^{\,'}R\right)\delta{}^{*}1,\label{eq:var L f}
\end{equation}
where we have been used $\delta f(R)=f^{\,'}\delta R$ with $f^{\,'}=\frac{df}{dR}$
. Consequently, the variation Eq.\eqref{eq:var L f} takes the following
form:

\begin{equation}
\mathcal{\delta L}_{f(R)}=f^{\,'}\mathcal{\delta L}_{EH}+\left(f-f^{\,'}R\right)\delta{}^{*}1.\label{tot-var}
\end{equation}
Inserting the total variation of E-H Lagrangian Eq.\eqref{EH-lag}
into Eq.\eqref{tot-var}, one obtains

\begin{eqnarray}
\delta\mathcal{L}_{f(R)} & = & f^{\,'}\left[D\delta\omega^{ab}\wedge{}^{*}e_{ab}+\delta e^{c}\wedge\left(R^{ab}\wedge{}^{*}e_{abc}\right)\right]+\left(f-f^{\,'}R\right)\delta e^{c}\wedge^{*}e_{c}\nonumber \\
 & = & f^{\,'}D\delta\omega^{ab}\wedge{}^{*}e_{ab}+\delta e^{c}\wedge\left[f^{\,'}R^{ab}\wedge{}^{*}e_{abc}+\left(f-f^{\,'}R\right){}^{*}e_{c}\right]\nonumber \\
 & = & \delta\omega^{ab}\wedge\left(df^{\,'}\land{}^{*}e_{ab}\right)+\delta e^{c}\wedge\left[f^{\,'}R^{ab}\wedge{}^{*}e_{abc}+\left(f-f^{\,'}R\right){}^{*}e_{c}\right].\label{eq:var Lf(R)}
\end{eqnarray}
In the last line we have used the fact that torsion is zero, i.e.,
$D\left(^{*}e_{ab}\right)=T^{c}\wedge{}^{*}e_{abc}=0$ and discharged
the boundary term and took $Df^{\,'}=df^{\,'}$. Since the connection
is a non-propagating field, one can solve variation of the connection
$\delta\omega^{ab}$ and express it in terms of the frame-like field
$\delta e^{a}$: 

\begin{eqnarray}
\delta\omega^{ab} & = & \frac{1}{2}i^{a}i^{b}\left(D\delta e^{c}\wedge e_{c}\right)-i^{a}(D\delta e^{b})+i^{b}(D\delta e^{a}).
\end{eqnarray}
Inserting this back in Eq.\eqref{eq:var Lf(R)}, one computes the
vierbein equation of motion in second-order formalism as 

\begin{equation}
f^{\,'}R^{ab}\wedge{}^{*}e_{abc}+(f-f^{\,'}R)^{*}e_{c}+2D\wedge{}^{*}(df^{\,'}\wedge e_{c})=0.\label{eq:eq of mot}
\end{equation}
This is the field equation for \textquotedbl metric-$f(R)$ theory\textquotedbl .
In terms of Einstein's tensor Eq.\eqref{eq:eq of mot} takes the following
form

\begin{equation}
f^{\,'}G_{\thinspace\thinspace c}^{a}{}^{*}e_{a}+\frac{1}{2}(f^{\,'}R-f)^{*}e_{c}-D{}^{*}(df^{\,'}\wedge e_{c})=0,\label{eq:f(R)-dif form}
\end{equation}
where Einstein tensor in flat space-time coordinates is given by $G_{\thinspace\thinspace c}^{a}=R_{\thinspace\thinspace c}^{a}-\frac{1}{2}\delta_{\thinspace\thinspace c}^{a}R$.
It is clearly seen here that the equation gives the well-known Einstein
field equations at the limit of $f(R)\rightarrow R$. In terms of
curved space-time coordinates Eq.\eqref{eq:f(R)-dif form} can be
expressed as Eq\eqref{f(R)-tensor}. A detailed calculation on $f(R)$
theory using differential forms can be found in\citep{Baykal:2010Unified,Baykal:2013MultiScalar}. 

\section{Maxwell extension of Einstein gravity}

In \citep{azcarraga2011generalized}, the authors presented the construction
of a local four-dimensional gauge theory based on the Maxwell algebra
and applied it to generalize Einstein's gravity (Maxwell Gravity).
Their construction of the action involves bilinear invariant curvature
2-forms and their action is not invariant under local Maxwell transformations
but only under local Lorentz transformations. The action respecting
the local Maxwell symmetry invariance is given in \citep{Cardenas:2022GeneralizedEinstein}.
They also considered bilinear invariant curvature 2-forms associated
with $AdSL_{4}$-valued 1-form gauge connection, and then constructed
a four-dimensional action that generalizes the Einstein gravity. They
showed that Maxwell Gravity can be obtained from $AdSL_{4}$-gravity
by means of the Inönü-Wigner contraction method\citep{Penafiel:2018AdS,Cardenas:2022GeneralizedEinstein}.
Both of these works make use of the bilinear invariant combination
of curvature 2-forms in their actions. On the contrary, we prefer
to make use of the linear combination of the gauge curvature 2-forms
that will be useful in the next section on the construction of the
Maxwell generalization of $f(R)$ theory.

We begin by giving an overview of the semi-simple extension of the
Poincare group \citep{soroka2012gauge}. The algebra of this group
was also re-derived in \citep{Gomis:2009Deformations} through the
deformation of Maxwell algebra. Nowadays, it is also known as the
AdS-Lorentz algebras \citep{durka2011gauged}. The $AdSL_{4}$ algebra,
on the other hand, is constructed from the $AdS$ algebra by means
of the S-expansion method introduced in \citep{Izaurieta:2006ExpandingLie,Salgado:2014soD1}.
The commutators of the algebra read

\begin{eqnarray}
\left[M_{ab},M_{cd}\right] & = & i\left(\eta_{ad}M_{bc}+\eta_{bc}M_{ad}-\eta_{ac}M_{bd}-\eta_{bd}M_{ac}\right),\nonumber \\
\left[M_{ab},Z_{cd}\right] & = & i\left(\eta_{ad}Z_{bc}+\eta_{bc}Z_{ad}-\eta_{ac}Z_{bd}-\eta_{bd}Z_{ac}\right),\nonumber \\
\left[Z_{ab},Z_{cd}\right] & = & i\mu\left(\eta_{ad}Z_{bc}+\eta_{bc}Z_{ad}-\eta_{ac}Z_{bd}-\eta_{bd}Z_{ac}\right),\nonumber \\
\left[P_{a},P_{b}\right] & = & i\lambda Z_{ab},\nonumber \\
\left[M_{ab},P_{c}\right] & = & i\left(\eta_{bc}P_{a}-\eta_{ac}P_{b}\right),\nonumber \\
\left[Z_{ab},P_{c}\right] & = & i\mu\left(\eta_{bc}P_{a}-\eta_{ac}P_{b}\right),
\end{eqnarray}
where the generators $X_{A}=\left\{ P_{a},M_{ab},Z_{ab}\right\} $
correspond to the translations, the Lorentz transformations and the
Maxwell symmetry. Here, for dimensional reasons the constant $\lambda$
is related to the cosmological constant $\Lambda$ and is given by
$\Lambda=\mu\lambda$ and the tangent space metric defined as $\eta_{ab}=\text{diag}\left(+,-,-,-\right)$.
On the contrary to Maxwell algebra, here $Z_{ab}$ are tensorial but
non-abelian generators. An interesting feature of the semi-simple
Poincare algebra is that it reproduces the Maxwell algebra through
the Inönü-Wigner contraction method.

In the following, we work along the lines of \citep{azcarraga2011generalized}
by using differential forms. To construct an action based on the semi-simple
Poincare algebra, we start from the following 1-form connection $A(x)=A^{A}X_{A}$:

\begin{equation}
A(x)=e^{a}P_{a}-\frac{1}{2}\omega^{ab}M_{ab}+\frac{1}{2}B^{ab}Z_{ab},\label{eq: gauge fields}
\end{equation}
where $A^{A}(x)=\left\{ e^{a},\omega^{ab},B^{ab}\right\} $ are the
gauge fields corresponding to the 16 generators of symmetry transformations.

The curvature 2-form $F=dA+\frac{i}{2}\left[A,A\right]$ associated
with the 1-form connection \eqref{eq: gauge fields} reads

\begin{eqnarray}
F\left(x\right) & = & F^{a}P_{a}-\frac{1}{2}R^{ab}M_{ab}+\frac{1}{2}F{}^{ab}Z_{ab},
\end{eqnarray}
where $F^{a}$ , $R^{ab}$ and $F^{ab}$ are generalized (Maxwell)
torsion, Ricci and Maxwell curvature 2-forms, respectively. Explicit
expressions for the curvature 2-forms are given by

\begin{eqnarray}
F^{a} & = & de^{a}+\omega_{\,\,c}^{a}\wedge e^{c}-\mu B_{\,\,c}^{a}\wedge e^{c}\nonumber \\
 & = & T^{a}-\mu B_{\,\,c}^{a}\wedge e^{c}\nonumber \\
 & = & \widetilde{D}e^{a},\label{eq:torsion Fa}
\end{eqnarray}

\begin{eqnarray}
R^{ab} & = & d\omega^{ab}+\omega_{\,\,c}^{a}\wedge\omega^{cb}=D\omega^{ab},
\end{eqnarray}

\begin{eqnarray}
F^{ab} & = & dB^{ab}+\omega_{\,\,c}^{[a}\wedge B^{c|b]}-\mu B_{\,\,c}^{a}\wedge B^{cb}-\lambda e^{a}\wedge e^{b}\nonumber \\
 & = & DB^{ab}-\mu B_{\,\,c}^{a}\wedge B^{cb}-\lambda e^{a}\wedge e^{b}\nonumber \\
 & = & \widetilde{D}B^{ab}-\lambda e^{a}\wedge e^{b},\label{eq:maxwell curv Fab}
\end{eqnarray}
where $D=d+\omega$ is the Lorentz exterior covariant derivative and
$\widetilde{D}=d+\widetilde{\omega}$ is the covariant exterior derivative
with respect to the shifted connection $\widetilde{\omega}^{ab}=\omega^{ab}-\mu B^{ab}$.

Within the scope of the study, we will make use of the shifted curvature
obtained by taking the difference between Ricci and Maxwell curvature
2-forms:

\begin{equation}
J^{ab}=R^{ab}(\omega)-\mu F^{ab}\equiv\widetilde{R}^{ab}(\widetilde{\omega})+\mu\lambda e^{a}\wedge e^{b}\label{eq:shifted cur}
\end{equation}
where $\widetilde{R}^{ab}(\widetilde{\omega})=\widetilde{D}\widetilde{\omega}^{ab}$
is the Ricci 2-form for the shifted connection 1-form $\widetilde{\omega}^{ab}=\omega^{ab}-\mu B^{ab}$.

Interestingly, the shifted connection $\widetilde{\omega}^{ab}$ can
be interpreted as an extension of the Riemannian connection $\omega^{ab}$
to a non-Riemannian one with torsion. In this sense, our result may
also be considered as the specific extension to a non-Riemannian framework
determined by the structure of the semi-simple Poincare algebra. Indeed,
in this context, the antisymmetry $B^{ab}=-B^{ba}$ implies that we
are dealing with an Einstein--Cartan geometry with non-metricity
tensor equal to zero because there is no symmetric part of the shifted
connection \citep{azcarraga2011generalized}.

In close analogy with the Einstein-Hilbert action, we start from the
following action of pure gravity:

\begin{equation}
S_{EHJ}=\frac{1}{2\kappa}\intop\mathcal{L}_{EHJ}
\end{equation}
If we promote $R^{ab}\rightarrow J^{ab}$, the Lagrangian 4-form in
differential form language will be given as

\begin{equation}
\mathcal{L}_{EHJ}=\frac{1}{2}\varepsilon_{abcd}J^{ab}\wedge e^{c}\wedge e^{d}=J^{ab}\wedge{}^{*}e_{ab}=J{}^{*}1.
\end{equation}
Clearly, the Lagrangian contains the Einstein--Hilbert, the cosmological
constant and a gravitational Maxwell term for the shifted connection.

At this point, there are two ways to proceed. First, one can either
vary the Lagrangian with respect to 1-form basis $e^{a}$ and the
connection 1-forms $\omega^{ab}$ independently (Palatini's method)
and then solve for the Maxwell torsion. Second, one can set the Maxwell
torsion equal to zero at the beginning and consider the zero Maxwell
torsion constrained variations of the Lagrangian. For the Einstein-Hilbert-like
Lagrangians (linear in curvatures) without any coupling to matter,
both of these cases imply the same set of field equations, namely
the source-free Einstein field equations with the cosmological term
\citep{Olmo:2011uz}. We derive the field equations by a first-order
formalism (Palatini's method). The dynamics of Maxwell gravity in
terms of vierbein $e^{a}$ and the shifted connection $\widetilde{\omega}^{ab}$
is derived by varying the Lagrangian:

\begin{eqnarray}
\delta\mathcal{L}_{EHJ} & = & \delta J^{ab}\wedge{}^{*}e_{ab}+J^{ab}\wedge\delta{}^{*}e_{ab}\nonumber \\
 & = & \left(\widetilde{D}\delta\widetilde{\omega}^{ab}+\mu\lambda\delta e^{[a}\wedge e^{|b]}\right)\wedge{}^{*}e_{ab}+J^{ab}\wedge\delta e^{c}\wedge{}^{*}e_{abc}\nonumber \\
 & = & \widetilde{D}\delta\widetilde{\omega}^{ab}\wedge{}^{*}e_{ab}+\delta e^{c}\wedge\left[J^{ab}\wedge{}^{*}e_{abc}+6\mu\lambda{}^{*}e_{c}\right].\label{eq:var ehj}
\end{eqnarray}
This full variation will be used for the Maxwell extension of $f(R)$
theory in the next section. One can write the Eq.\eqref{eq:var ehj}
up to a total exterior derivative term and set the variation to zero: 

\begin{equation}
\delta\mathcal{L}_{EHJ}=\delta\widetilde{\omega}^{ab}\wedge F^{c}\wedge{}^{*}e_{abc}+\delta e^{c}\wedge\left[J^{ab}\wedge{}^{*}e_{abc}+6\mu\lambda{}^{*}e_{c}\right]=0,
\end{equation}
gives two equations of motion. One is the torsion equation,

\begin{equation}
F^{c}=0\Rightarrow T^{c}=B_{\thinspace d}^{c}\wedge e^{d}.\label{eq:tor-B}
\end{equation}
It is worth stressing that torsion comes quite naturally since it
is introduced by gauge fields $B^{ab}$. Although the torsion associated
with the gauge field $\omega^{ab}$ need not be zero, the torsion
of the re-defined fields $\widetilde{\omega}^{ab}$ should be zero.
In the general case, from equation \eqref{eq:tor-B} it is seen that
$B$-field is the source of torsion, and hence the connection is no
longer on-shell torsion-free even in the source-free region.

The other is the generalized Einstein's equation:

\begin{equation}
J^{ab}\wedge{}^{*}e_{abc}+6\mu\lambda{}^{*}e_{c}=0\Rightarrow R^{ab}\wedge{}^{*}e_{abc}+6\mu\lambda{}^{*}e_{c}=\mu F^{ab}\wedge{}^{*}e_{abc}.
\end{equation}
and it can be decomposed as

\begin{equation}
R^{ab}\wedge{}^{*}e_{abc}+6\mu\lambda{}^{*}e_{c}=\mu F^{ab}\wedge{}^{*}e_{abc}.\label{eq:gen-eins}
\end{equation}
To recast it in a more familiar form one can use the relations: $e_{a}^{\mu}e_{b}^{\nu}R^{ab}=\frac{1}{2}R{}_{\thinspace\thinspace\thinspace\thinspace\rho\sigma}^{\mu\nu}dx^{\rho}\wedge dx^{\sigma}$,
$e_{a}^{\mu}e_{b}^{\nu}F^{ab}=\frac{1}{2}F_{\thinspace\thinspace\thinspace\thinspace\rho\sigma}^{\mu\nu}dx^{\rho}\wedge dx^{\sigma}$

\begin{equation}
R_{\,\,\sigma}^{\nu}-\frac{1}{2}R\delta_{\,\,\sigma}^{\nu}-6\mu\lambda\delta_{\,\,\sigma}^{\nu}=\mu T_{\,\,\sigma}^{\nu}\left(B\right),\label{eq:gen-eins-tensor}
\end{equation}
where
\begin{equation}
T_{\,\,\sigma}^{\nu}\left(B\right)=e_{a}^{\mu}e_{b}^{\nu}\left(D_{[\mu}B_{\sigma]}^{ab}-\mu B_{[\mu\,\,c}^{a}\wedge B_{\sigma]}^{cb}\right)-\frac{1}{2}\delta_{\,\,\sigma}^{\nu}e_{a}^{\gamma}e_{b}^{\kappa}\left(D_{[\gamma}B_{\kappa]}^{ab}-\mu B_{[\gamma\,\,c}^{a}\wedge B_{\kappa]}^{cb}\right),
\end{equation}
is the energy-momentum tensor for the background field and the square
brackets around the indices imply antisymmetrization. It is important
to note that the cosmological constant term is naturally contained
here without the need to introduce it by hand. 

\section{Maxwell extension of $f\left(R\right)$ gravity}

We propose a new version of $f(R)$ theory by combining two existing
ideas, i.e., by invoking the gauge theory of gravity as the proper
description of gravitational effects and by assuming semi-simple extended
Poincare symmetry as the underlying gauge group of the universe. Now,
to construct a Maxwell generalized $f(R)$ action from the curvature
of the semi-simple Poincare algebra, we consider the shifted curvature
2-form $J^{ab}$ in Eq.\eqref{eq:shifted cur}. If we promote $f(R)\rightarrow f(J)$,
the Lagrangian 4-form for the Maxwell generalized $f(R)$ theory can
be written in differential form language as

\begin{equation}
\mathcal{L}_{f(J)}=f(J)^{*}1.
\end{equation}
Following the same procedure presented in Section 2, the total variation
of the 4-form $\mathcal{L}_{f(J)}$ can be calculated as follows.
Using again the product rule, the variation takes the following form

\begin{equation}
\mathcal{\delta L}_{f(J)}=f^{\,'}\mathcal{\delta L}_{EHJ}+\left(f-f^{\,'}J\right)\delta{}^{*}1,\label{eq:deltaLfJ}
\end{equation}
 where this time $\delta f(J)=f^{\,'}\delta J$ with $f^{\,'}=\frac{df}{dJ}$
. 

Inserting the variation of $\mathcal{\delta L}_{EHJ}$ from Eq.\eqref{eq:var ehj}
into Eq.\eqref{eq:deltaLfJ}, one gets

\begin{equation}
\mathcal{\delta L}_{f(J)}=f^{\,'}\left[\widetilde{D}\delta\widetilde{\omega}^{ab}\wedge{}^{*}e_{ab}+\delta e^{c}\wedge\left(J^{ab}\wedge{}^{*}e_{abc}+6\mu\lambda{}^{*}e_{c}\right)\right]+\left(f-f^{\,'}J\right)\delta e^{c}\wedge^{*}e_{c}.
\end{equation}
 Rearranging gives: 
\begin{eqnarray}
\delta\mathcal{L}_{f(J)} & = & f^{\,'}\widetilde{D}\delta\widetilde{\omega}^{ab}\wedge{}^{*}e_{ab}+\delta e^{c}\wedge\left[f^{\,'}J^{ab}\wedge{}^{*}e_{abc}+\left(f-f^{\,'}J+6\mu\lambda f^{\,'}\right){}^{*}e_{c}\right]\nonumber \\
 & = & \delta\widetilde{\omega}^{ab}\wedge\widetilde{D}\left(f^{\,'}{}^{*}e_{ab}\right)+\delta e^{c}\wedge\left[f^{\,'}J^{ab}\wedge{}^{*}e_{abc}+\left(f-f^{\,'}J+6\mu\lambda f^{\,'}\right){}^{*}e_{c}\right]\nonumber \\
 & = & \delta\widetilde{\omega}^{ab}\wedge\left(df^{\,'}\land{}^{*}e_{ab}\right)+\delta e^{c}\wedge\left[f^{\,'}J^{ab}\wedge{}^{*}e_{abc}+\left(f-f^{\,'}J+6\mu\lambda f^{\,'}\right){}^{*}e_{c}\right].\label{eq:delta var f J-2}
\end{eqnarray}

In the last line we have used the fact that Maxwell-torsion is zero,
i.e., $\widetilde{D}\left(^{*}e_{ab}\right)=F^{c}\wedge{}^{*}e_{abc}=0$
and discharged the boundary term and took $\widetilde{D}f^{\,'}=df^{\,'}$.
It is helpful to remark here that the $f(R)$-theories of gravity
with torsion is also considered in \citep{Capozziello:2007f(R)gravity}
where torsion is geometrically inserted in f(R) gravity and in \citep{Capozziello:2008f(R)gravity}
where $f(R)$ gravity with torsion is formulated in the $J$-bundle
approach. In \citep{Capozziello:2007f(R)gravity}, the authors start
with the metric-affine formulation in which the affine connection
is assumed to be metric compatible. As is well known for a given metric
tensor $g_{\mu\nu},$ every metric connection is expressible as the
sum of Levi-Civita connection and contorsion tensor, i.e., $\Gamma_{\alpha\beta}^{\mu}=\mathring{\Gamma}_{\alpha\beta}^{\mu}+K_{\alpha\beta}^{\mu}$.
The antisymmetry property of the contorsion tensor $K_{\alpha\beta}^{\mu}=-K_{\beta\alpha}^{\mu}$
ensures the metric compatibility of the connection $\Gamma_{\alpha\beta}^{\mu}$.
In this way, one can identify the actual degrees of freedom of the
theory with the independent components of the metric $g_{\mu\nu}$
and the contorsion tensor $K_{\alpha\beta}^{\mu}$. In \citep{Capozziello:2008f(R)gravity},
the authors used the $J$-bundle approach to construct the $f(R)$-theories
of gravity with torsion. In the $J$-bundles framework, one starts
from several Lagrangians densities which depend on the fields only
through their antisymmetric combinations. This is the case of the
Einstein--Hilbert like Lagrangian which, in the tetrad-affine formulation,
depends on the antisymmetric derivatives of the spin-connection through
the curvature. In view of this fact, they defined a suitable quotient
space of the first jet bundle, made equivalent two sections which
have a first order contact with respect to the exterior differentiation,
instead of the whole set of derivatives. The resulting fiber coordinates
of the so defined new spaces are exactly the antisymmetric combinations
appearing in the Lagrangian densities. For general relativity $(GR)$
as well as $f(R)$ gravity, they also showed that the fiber coordinates
of the quotient space can be identified with the components of the
torsion and curvature tensors. For further information, we refer the
reader to \citep{Capozziello:2007f(R)gravity,Capozziello:2008f(R)gravity}
and the references therein.

The very fact that both approaches, being formally different, give
essentially equivalent results and in both approaches a torsion arises
from the non-linearity of the gravitational Lagrangian function even
in the absence of spin and its existence does not affect the metric
field equations. Our approach, on the other hand, starts from the
antisymmetric shifted connection $\widetilde{\omega}^{ab}=\omega^{ab}-\mu B^{ab}$
in the tetrad-affine formulation, and therefore the antisymmetric
background field $B^{ab}$ is the source of the torsion.

After these remarks were given, we proceed as in the previous section.
The connection is again a non-propagating field, then we can define
a relation between vierbein $\delta e^{a}$ and shifted connection
$\delta\widetilde{\omega}^{ab}$ variations for the zero-Maxwell torsion
case as follows:

\begin{eqnarray}
\delta\widetilde{\omega}^{ab} & = & \frac{1}{2}i^{a}i^{b}\left(\widetilde{D}\delta e^{c}\wedge e_{c}\right)-i^{a}(\widetilde{D}\delta e^{b})+i^{b}(\widetilde{D}\delta e^{a}).
\end{eqnarray}
Inserting this in Eq.\eqref{eq:delta var f J-2}, one can obtain the
field equations of motion in second-order formalism as 

\begin{equation}
f^{\,'}J^{ab}\wedge{}^{*}e_{abc}+(f-f^{\,'}J+6\mu\lambda f^{\,'})^{*}e_{c}+2\widetilde{D}\wedge{}^{*}(df^{\,'}\wedge e_{c})=0.\label{eqofmot1}
\end{equation}

The Maxwell modified $f(R)$-metric field equation for the presence
of the background field $B^{ab}$ that follows from Eq.\eqref{eqofmot1}
takes the form

\begin{equation}
f^{\,'}(G_{\thinspace\thinspace c}^{a}-3\mu\lambda\delta_{\thinspace\thinspace c}^{a}){}^{*}e_{a}-\frac{1}{2}(f-f^{\,'}R)^{*}e_{c}-\widetilde{D}\wedge{}^{*}(df^{\,'}\wedge e_{c})=\mu f^{\,'}F_{\thinspace\thinspace c}^{a}\wedge{}^{*}e_{a},
\end{equation}
where $G_{\thinspace\thinspace c}^{a}=R_{\thinspace\thinspace c}^{a}-\frac{1}{2}\delta_{\thinspace\thinspace c}^{a}R$
is the Einstein tensor in flat space-time coordinates. Maxwell extended
$f(R)$ gravity with cosmological constant are naturally contained
here without the need to introduce a cosmological constant by hand.
To recast it in a more familiar form one can pass from the tangent
indices to world indices by using the relations: $e_{a}^{\mu}e_{b}^{\nu}R^{ab}=\frac{1}{2}R{}_{\thinspace\thinspace\thinspace\thinspace\rho\sigma}^{\mu\nu}dx^{\rho}\wedge dx^{\sigma}$,
$e_{a}^{\mu}e_{b}^{\nu}F^{ab}=\frac{1}{2}F_{\thinspace\thinspace\thinspace\thinspace\rho\sigma}^{\mu\nu}dx^{\rho}\wedge dx^{\sigma}$,
one gets the following field equation,

\begin{equation}
R_{\,\,\sigma}^{\nu}-\frac{1}{2}R\delta_{\,\,\sigma}^{\nu}-6\mu\lambda\delta_{\,\,\sigma}^{\nu}=T_{\,\,\sigma}^{\nu}\left(eff\right)+\mu T_{\,\,\sigma}^{\nu}\left(B\right),\label{eq:s1}
\end{equation}
where

\begin{equation}
T_{\,\,\sigma}^{\nu}\left(eff\right)=\frac{1}{f^{\,'}}\left\{ \nabla^{\nu}\nabla_{\sigma}f^{\,'}-\delta_{\thinspace\thinspace\sigma}^{\nu}\Delta f^{\,'}+\frac{1}{2}\delta_{\thinspace\thinspace\sigma}^{\nu}\left(f-f^{\,'}R\right)\right\} ,\label{eq:s2}
\end{equation}
and

\begin{equation}
T_{\,\,\sigma}^{\nu}\left(B\right)=e_{a}^{\mu}e_{b}^{\nu}\left(D_{[\mu}B_{\sigma]}^{ab}-\mu B_{[\mu\,\,c}^{a}\wedge B_{\sigma]}^{cb}\right)-\frac{1}{2}\delta_{\,\,\sigma}^{\nu}e_{a}^{\gamma}e_{b}^{\kappa}\left(D_{[\gamma}B_{\kappa]}^{ab}-\mu B_{[\gamma\,\,c}^{a}\wedge B_{\kappa]}^{cb}\right),\label{eq:s3}
\end{equation}
here $T_{\,\,\sigma}^{\nu}\left(eff\right)$ is an extra gravitational
energy-momentum tensor due to higher order curvature effects and $T_{\,\,\rho}^{\mu}\left(B\right)$
is energy-momentum tensor for the background gauge field $B^{ab}(x)$
due to Maxwell extension. As is expected, in the limit $\mu\rightarrow0$,
Eqs.\eqref{eq:s1}, \eqref{eq:s2} and \eqref{eq:s3} turn to the
well known equations of motion for metric-$f(R)$ gravity. Result
obtained in \eqref{eq:s1} is actually $AdSL_{4}$ extension of $f(R)$
gravity. If one performs a Inönü-Wigner contraction of $AdSL_{4}$
down to Maxwell group by re-scaling the generators $P_{a}\rightarrow\xi P_{a},Z_{ab}\rightarrow\xi^{2}Z_{ab}$
as well as the gauge fields $e^{a}\rightarrow\xi^{-1}e^{a},B^{ab}\rightarrow\xi^{-2}B^{ab}$
and then taking the limit $\xi\rightarrow\infty$, we obtain the equations
of motion for the Maxwell extension of $f(R)$ gravity with the cosmological
term. In doing so, the derivative in Eq.\eqref{eq:s2} with respect
to $J$ becomes derivative with respect to $R$ and the last energy-momentum
tensor changes to

\begin{equation}
T_{\,\,\sigma}^{\nu}\left(B\right)=e_{a}^{\mu}e_{b}^{\nu}D_{[\mu}B_{\sigma]}^{ab}-\frac{1}{2}\delta_{\,\,\sigma}^{\nu}e_{a}^{\gamma}e_{b}^{\kappa}D_{[\gamma}B_{\kappa]}^{ab}.\label{eq:s3-1}
\end{equation}

\section{Conclusion}

In this paper, we have presented the Maxwell extension of $f(R)$
theory of gravity as the gravitational part of the action is a function
of the shifted curvature scalar $J$, i.e., $f(J)$. This could be
a linear function, or non-linear. We have considered the curvature
2-forms associated with the semi-simple extension of Poincare algebra
$(AdSL_{4})$-valued one-form gauge connection, and then we constructed
a four-dimensional action that generalizes the $f(R)$ gravity. It
is shown that the Maxwell extension of $f(R)$ gravity can be obtained
from $AdSL_{4}$-$f(R)$ gravity making use of the Inönü-Wigner contraction
method. It is found that the Maxwell extension modifies the results
of the metric-$f(R)$ gravity not only by introducing the cosmological
constant term but also the new gauge fields $B_{\lyxmathsym{\textmu}}^{ab}(x$)
terms. These could play the role of inflaton vector fields which drive
accelerated expansion from Maxwell $f(R)$ gravity \citep{azcarraga2011generalized,Azcarraga2013maxwellApplication}.
It would be interesting to apply our formalism to Gauss-Bonnet type
generalization of $f(R)$ gravity respecting the local Maxwell symmetry.
In such a theory, the shifted curvature, Maxwell torsion and the vector
gauge fields may give rise to a modified $f(R)$ gravity theory which
is capable, in principle, to address the problem of the dark side
of the universe in a very general geometric scheme. 
\begin{acknowledgments}
This work was supported by the Scientific and Technological Research
Council of Turkey (TÜB\.{I}TAK) Research project No. 118F364. 
\end{acknowledgments}

\bibliographystyle{apsrev4-2}
\phantomsection\addcontentsline{toc}{section}{\refname}\bibliography{Maxwell_FR}

\begin{thebibliography}{51}%
\makeatletter
\providecommand \@ifxundefined [1]{%
 \@ifx{#1\undefined}
}%
\providecommand \@ifnum [1]{%
 \ifnum #1\expandafter \@firstoftwo
 \else \expandafter \@secondoftwo
 \fi
}%
\providecommand \@ifx [1]{%
 \ifx #1\expandafter \@firstoftwo
 \else \expandafter \@secondoftwo
 \fi
}%
\providecommand \natexlab [1]{#1}%
\providecommand \enquote  [1]{``#1''}%
\providecommand \bibnamefont  [1]{#1}%
\providecommand \bibfnamefont [1]{#1}%
\providecommand \citenamefont [1]{#1}%
\providecommand \href@noop [0]{\@secondoftwo}%
\providecommand \href [0]{\begingroup \@sanitize@url \@href}%
\providecommand \@href[1]{\@@startlink{#1}\@@href}%
\providecommand \@@href[1]{\endgroup#1\@@endlink}%
\providecommand \@sanitize@url [0]{\catcode `\\12\catcode `\$12\catcode
  `\&12\catcode `\#12\catcode `\^12\catcode `\_12\catcode `\%12\relax}%
\providecommand \@@startlink[1]{}%
\providecommand \@@endlink[0]{}%
\providecommand \url  [0]{\begingroup\@sanitize@url \@url }%
\providecommand \@url [1]{\endgroup\@href {#1}{\urlprefix }}%
\providecommand \urlprefix  [0]{URL }%
\providecommand \Eprint [0]{\href }%
\providecommand \doibase [0]{https://doi.org/}%
\providecommand \selectlanguage [0]{\@gobble}%
\providecommand \bibinfo  [0]{\@secondoftwo}%
\providecommand \bibfield  [0]{\@secondoftwo}%
\providecommand \translation [1]{[#1]}%
\providecommand \BibitemOpen [0]{}%
\providecommand \bibitemStop [0]{}%
\providecommand \bibitemNoStop [0]{.\EOS\space}%
\providecommand \EOS [0]{\spacefactor3000\relax}%
\providecommand \BibitemShut  [1]{\csname bibitem#1\endcsname}%
\let\auto@bib@innerbib\@empty
\bibitem [{\citenamefont {Riess}\ \emph {et~al.}(1998)\citenamefont {Riess}
  \emph {et~al.}}]{riess1998observational}%
  \BibitemOpen
  \bibfield  {author} {\bibinfo {author} {\bibfnamefont {A.~G.}\ \bibnamefont
  {Riess}} \emph {et~al.} (\bibinfo {collaboration} {Supernova Search Team}),\
  }\href {https://doi.org/10.1086/300499} {\bibfield  {journal} {\bibinfo
  {journal} {Astron. J.}\ }\textbf {\bibinfo {volume} {116}},\ \bibinfo {pages}
  {1009} (\bibinfo {year} {1998})}\BibitemShut {NoStop}%
\bibitem [{\citenamefont {Trimble}(1987)}]{Trimble:1987Existence}%
  \BibitemOpen
  \bibfield  {author} {\bibinfo {author} {\bibfnamefont {V.}~\bibnamefont
  {Trimble}},\ }\href {https://doi.org/10.1146/annurev.aa.25.090187.002233}
  {\bibfield  {journal} {\bibinfo  {journal} {Ann. Rev. Astron. Astrophys.}\
  }\textbf {\bibinfo {volume} {25}},\ \bibinfo {pages} {425} (\bibinfo {year}
  {1987})}\BibitemShut {NoStop}%
\bibitem [{\citenamefont {Padmanabhan}(2003)}]{Padmanabhan:2003Cosmological}%
  \BibitemOpen
  \bibfield  {author} {\bibinfo {author} {\bibfnamefont {T.}~\bibnamefont
  {Padmanabhan}},\ }\href {https://doi.org/10.1016/S0370-1573(03)00120-0}
  {\bibfield  {journal} {\bibinfo  {journal} {Phys. Rept.}\ }\textbf {\bibinfo
  {volume} {380}},\ \bibinfo {pages} {235} (\bibinfo {year}
  {2003})}\BibitemShut {NoStop}%
\bibitem [{\citenamefont {Peebles}\ and\ \citenamefont
  {Ratra}(2003)}]{Peebles:2003Cosmological}%
  \BibitemOpen
  \bibfield  {author} {\bibinfo {author} {\bibfnamefont {P.~J.~E.}\
  \bibnamefont {Peebles}}\ and\ \bibinfo {author} {\bibfnamefont
  {B.}~\bibnamefont {Ratra}},\ }\href
  {https://doi.org/10.1103/RevModPhys.75.559} {\bibfield  {journal} {\bibinfo
  {journal} {Rev. Mod. Phys.}\ }\textbf {\bibinfo {volume} {75}},\ \bibinfo
  {pages} {559} (\bibinfo {year} {2003})}\BibitemShut {NoStop}%
\bibitem [{\citenamefont {Buchdahl}(1970)}]{Buchdahl:1970Non}%
  \BibitemOpen
  \bibfield  {author} {\bibinfo {author} {\bibfnamefont {H.~A.}\ \bibnamefont
  {Buchdahl}},\ }\href@noop {} {\bibfield  {journal} {\bibinfo  {journal} {Mon.
  Not. Roy. Astron. Soc.}\ }\textbf {\bibinfo {volume} {150}},\ \bibinfo
  {pages} {1} (\bibinfo {year} {1970})}\BibitemShut {NoStop}%
\bibitem [{\citenamefont {Starobinsky}(1980)}]{Starobinsky:1980ANew}%
  \BibitemOpen
  \bibfield  {author} {\bibinfo {author} {\bibfnamefont {A.~A.}\ \bibnamefont
  {Starobinsky}},\ }\href {https://doi.org/10.1016/0370-2693(80)90670-X}
  {\bibfield  {journal} {\bibinfo  {journal} {Phys. Lett. B}\ }\textbf
  {\bibinfo {volume} {91}},\ \bibinfo {pages} {99} (\bibinfo {year}
  {1980})}\BibitemShut {NoStop}%
\bibitem [{\citenamefont {Nojiri}\ and\ \citenamefont
  {Odintsov}(2003)}]{nojiri2003modified}%
  \BibitemOpen
  \bibfield  {author} {\bibinfo {author} {\bibfnamefont {S.}~\bibnamefont
  {Nojiri}}\ and\ \bibinfo {author} {\bibfnamefont {S.~D.}\ \bibnamefont
  {Odintsov}},\ }\href {https://doi.org/10.1103/PhysRevD.68.123512} {\bibfield
  {journal} {\bibinfo  {journal} {Phys. Rev. D}\ }\textbf {\bibinfo {volume}
  {68}},\ \bibinfo {pages} {123512} (\bibinfo {year} {2003})}\BibitemShut
  {NoStop}%
\bibitem [{\citenamefont {Capozziello}\ \emph {et~al.}(2003)\citenamefont
  {Capozziello}, \citenamefont {Cardone}, \citenamefont {Carloni},\ and\
  \citenamefont {Troisi}}]{Capozziello:2003quintessence}%
  \BibitemOpen
  \bibfield  {author} {\bibinfo {author} {\bibfnamefont {S.}~\bibnamefont
  {Capozziello}}, \bibinfo {author} {\bibfnamefont {V.~F.}\ \bibnamefont
  {Cardone}}, \bibinfo {author} {\bibfnamefont {S.}~\bibnamefont {Carloni}},\
  and\ \bibinfo {author} {\bibfnamefont {A.}~\bibnamefont {Troisi}},\ }\href
  {https://doi.org/10.1142/S0218271803004407} {\bibfield  {journal} {\bibinfo
  {journal} {Int. J. Mod. Phys. D}\ }\textbf {\bibinfo {volume} {12}},\
  \bibinfo {pages} {1969} (\bibinfo {year} {2003})}\BibitemShut {NoStop}%
\bibitem [{\citenamefont {Carroll}\ \emph {et~al.}(2004)\citenamefont
  {Carroll}, \citenamefont {Duvvuri}, \citenamefont {Trodden},\ and\
  \citenamefont {Turner}}]{Carroll:2004IsCosmicSpeed}%
  \BibitemOpen
  \bibfield  {author} {\bibinfo {author} {\bibfnamefont {S.~M.}\ \bibnamefont
  {Carroll}}, \bibinfo {author} {\bibfnamefont {V.}~\bibnamefont {Duvvuri}},
  \bibinfo {author} {\bibfnamefont {M.}~\bibnamefont {Trodden}},\ and\ \bibinfo
  {author} {\bibfnamefont {M.~S.}\ \bibnamefont {Turner}},\ }\href
  {https://doi.org/10.1103/PhysRevD.70.043528} {\bibfield  {journal} {\bibinfo
  {journal} {Phys. Rev. D}\ }\textbf {\bibinfo {volume} {70}},\ \bibinfo
  {pages} {043528} (\bibinfo {year} {2004})}\BibitemShut {NoStop}%
\bibitem [{\citenamefont {Nojiri}\ and\ \citenamefont
  {Odintsov}(2007)}]{nojiri2007introduction}%
  \BibitemOpen
  \bibfield  {author} {\bibinfo {author} {\bibfnamefont {S.}~\bibnamefont
  {Nojiri}}\ and\ \bibinfo {author} {\bibfnamefont {S.~D.}\ \bibnamefont
  {Odintsov}},\ }\href {https://doi.org/10.1142/S0219887807001928} {\bibfield
  {journal} {\bibinfo  {journal} {Int. J. Geom. Methods Mod. Phys.}\ }\textbf
  {\bibinfo {volume} {4}},\ \bibinfo {pages} {115} (\bibinfo {year}
  {2007})}\BibitemShut {NoStop}%
\bibitem [{\citenamefont {Hu}\ and\ \citenamefont
  {Sawicki}(2007)}]{Hu:2007ModelsOfF(R)}%
  \BibitemOpen
  \bibfield  {author} {\bibinfo {author} {\bibfnamefont {W.}~\bibnamefont
  {Hu}}\ and\ \bibinfo {author} {\bibfnamefont {I.}~\bibnamefont {Sawicki}},\
  }\href {https://doi.org/10.1103/PhysRevD.76.064004} {\bibfield  {journal}
  {\bibinfo  {journal} {Phys. Rev. D}\ }\textbf {\bibinfo {volume} {76}},\
  \bibinfo {pages} {064004} (\bibinfo {year} {2007})}\BibitemShut {NoStop}%
\bibitem [{\citenamefont {Appleby}\ and\ \citenamefont
  {Battye}(2007)}]{Appleby:2007DoConsistent}%
  \BibitemOpen
  \bibfield  {author} {\bibinfo {author} {\bibfnamefont {S.~A.}\ \bibnamefont
  {Appleby}}\ and\ \bibinfo {author} {\bibfnamefont {R.~A.}\ \bibnamefont
  {Battye}},\ }\href {https://doi.org/10.1016/j.physletb.2007.08.037}
  {\bibfield  {journal} {\bibinfo  {journal} {Phys. Lett. B}\ }\textbf
  {\bibinfo {volume} {654}},\ \bibinfo {pages} {7} (\bibinfo {year}
  {2007})}\BibitemShut {NoStop}%
\bibitem [{\citenamefont
  {Starobinsky}(2007)}]{Starobinsky:2007DisappearingCCF(R)}%
  \BibitemOpen
  \bibfield  {author} {\bibinfo {author} {\bibfnamefont {A.~A.}\ \bibnamefont
  {Starobinsky}},\ }\href {https://doi.org/10.1134/S0021364007150027}
  {\bibfield  {journal} {\bibinfo  {journal} {JETP Lett.}\ }\textbf {\bibinfo
  {volume} {86}},\ \bibinfo {pages} {157} (\bibinfo {year} {2007})}\BibitemShut
  {NoStop}%
\bibitem [{\citenamefont {Amendola}\ \emph {et~al.}(2007)\citenamefont
  {Amendola}, \citenamefont {Gannouji}, \citenamefont {Polarski},\ and\
  \citenamefont {Tsujikawa}}]{Amendola:2007Conditions}%
  \BibitemOpen
  \bibfield  {author} {\bibinfo {author} {\bibfnamefont {L.}~\bibnamefont
  {Amendola}}, \bibinfo {author} {\bibfnamefont {R.}~\bibnamefont {Gannouji}},
  \bibinfo {author} {\bibfnamefont {D.}~\bibnamefont {Polarski}},\ and\
  \bibinfo {author} {\bibfnamefont {S.}~\bibnamefont {Tsujikawa}},\ }\href
  {https://doi.org/10.1103/PhysRevD.75.083504} {\bibfield  {journal} {\bibinfo
  {journal} {Phys. Rev. D}\ }\textbf {\bibinfo {volume} {75}},\ \bibinfo
  {pages} {083504} (\bibinfo {year} {2007})}\BibitemShut {NoStop}%
\bibitem [{\citenamefont {Sotiriou}\ and\ \citenamefont
  {Liberati}(2007)}]{Sotiriou:2007MetricAffine}%
  \BibitemOpen
  \bibfield  {author} {\bibinfo {author} {\bibfnamefont {T.~P.}\ \bibnamefont
  {Sotiriou}}\ and\ \bibinfo {author} {\bibfnamefont {S.}~\bibnamefont
  {Liberati}},\ }\href {https://doi.org/10.1016/j.aop.2006.06.002} {\bibfield
  {journal} {\bibinfo  {journal} {Annals Phys.}\ }\textbf {\bibinfo {volume}
  {322}},\ \bibinfo {pages} {935} (\bibinfo {year} {2007})}\BibitemShut
  {NoStop}%
\bibitem [{\citenamefont {Sotiriou}\ and\ \citenamefont
  {Faraoni}(2010)}]{Sotiriou:2010FR}%
  \BibitemOpen
  \bibfield  {author} {\bibinfo {author} {\bibfnamefont {T.~P.}\ \bibnamefont
  {Sotiriou}}\ and\ \bibinfo {author} {\bibfnamefont {V.}~\bibnamefont
  {Faraoni}},\ }\href {https://doi.org/10.1103/RevModPhys.82.451} {\bibfield
  {journal} {\bibinfo  {journal} {Rev. Mod. Phys.}\ }\textbf {\bibinfo {volume}
  {82}},\ \bibinfo {pages} {451} (\bibinfo {year} {2010})}\BibitemShut
  {NoStop}%
\bibitem [{\citenamefont {Nojiri}\ and\ \citenamefont
  {Odintsov}(2011)}]{nojiri2011unified}%
  \BibitemOpen
  \bibfield  {author} {\bibinfo {author} {\bibfnamefont {S.}~\bibnamefont
  {Nojiri}}\ and\ \bibinfo {author} {\bibfnamefont {S.~D.}\ \bibnamefont
  {Odintsov}},\ }\href {https://doi.org/10.1016/j.physrep.2011.04.001}
  {\bibfield  {journal} {\bibinfo  {journal} {Phys. Rept.}\ }\textbf {\bibinfo
  {volume} {505}},\ \bibinfo {pages} {59} (\bibinfo {year} {2011})}\BibitemShut
  {NoStop}%
\bibitem [{\citenamefont {De~Felice}\ and\ \citenamefont
  {Tsujikawa}(2010)}]{DeFelice:2010FR}%
  \BibitemOpen
  \bibfield  {author} {\bibinfo {author} {\bibfnamefont {A.}~\bibnamefont
  {De~Felice}}\ and\ \bibinfo {author} {\bibfnamefont {S.}~\bibnamefont
  {Tsujikawa}},\ }\href {https://doi.org/10.12942/lrr-2010-3} {\bibfield
  {journal} {\bibinfo  {journal} {Living Rev. Rel.}\ }\textbf {\bibinfo
  {volume} {13}},\ \bibinfo {pages} {3} (\bibinfo {year} {2010})}\BibitemShut
  {NoStop}%
\bibitem [{\citenamefont {Capozziello}\ and\ \citenamefont
  {De~Laurentis}(2011)}]{Capozziello:2011Extended}%
  \BibitemOpen
  \bibfield  {author} {\bibinfo {author} {\bibfnamefont {S.}~\bibnamefont
  {Capozziello}}\ and\ \bibinfo {author} {\bibfnamefont {M.}~\bibnamefont
  {De~Laurentis}},\ }\href {https://doi.org/10.1016/j.physrep.2011.09.003}
  {\bibfield  {journal} {\bibinfo  {journal} {Phys. Rept.}\ }\textbf {\bibinfo
  {volume} {509}},\ \bibinfo {pages} {167} (\bibinfo {year}
  {2011})}\BibitemShut {NoStop}%
\bibitem [{\citenamefont {Clifton}\ \emph {et~al.}(2012)\citenamefont
  {Clifton}, \citenamefont {Ferreira}, \citenamefont {Padilla},\ and\
  \citenamefont {Skordis}}]{Clifton:2011Modified}%
  \BibitemOpen
  \bibfield  {author} {\bibinfo {author} {\bibfnamefont {T.}~\bibnamefont
  {Clifton}}, \bibinfo {author} {\bibfnamefont {P.~G.}\ \bibnamefont
  {Ferreira}}, \bibinfo {author} {\bibfnamefont {A.}~\bibnamefont {Padilla}},\
  and\ \bibinfo {author} {\bibfnamefont {C.}~\bibnamefont {Skordis}},\ }\href
  {https://doi.org/10.1016/j.physrep.2012.01.001} {\bibfield  {journal}
  {\bibinfo  {journal} {Phys. Rept.}\ }\textbf {\bibinfo {volume} {513}},\
  \bibinfo {pages} {1} (\bibinfo {year} {2012})}\BibitemShut {NoStop}%
\bibitem [{\citenamefont {Olmo}(2011)}]{Olmo:2011uz}%
  \BibitemOpen
  \bibfield  {author} {\bibinfo {author} {\bibfnamefont {G.~J.}\ \bibnamefont
  {Olmo}},\ }\href {https://doi.org/10.1142/S0218271811018925} {\bibfield
  {journal} {\bibinfo  {journal} {Int. J. Mod. Phys. D}\ }\textbf {\bibinfo
  {volume} {20}},\ \bibinfo {pages} {413} (\bibinfo {year} {2011})}\BibitemShut
  {NoStop}%
\bibitem [{\citenamefont {Nojiri}\ \emph {et~al.}(2017)\citenamefont {Nojiri},
  \citenamefont {Odintsov},\ and\ \citenamefont
  {Oikonomou}}]{Nojiri:2017Modified}%
  \BibitemOpen
  \bibfield  {author} {\bibinfo {author} {\bibfnamefont {S.}~\bibnamefont
  {Nojiri}}, \bibinfo {author} {\bibfnamefont {S.~D.}\ \bibnamefont
  {Odintsov}},\ and\ \bibinfo {author} {\bibfnamefont {V.~K.}\ \bibnamefont
  {Oikonomou}},\ }\href {https://doi.org/10.1016/j.physrep.2017.06.001}
  {\bibfield  {journal} {\bibinfo  {journal} {Phys. Rept.}\ }\textbf {\bibinfo
  {volume} {692}},\ \bibinfo {pages} {1} (\bibinfo {year} {2017})}\BibitemShut
  {NoStop}%
\bibitem [{\citenamefont {Heisenberg}(2019)}]{Heisenberg:2019systematic}%
  \BibitemOpen
  \bibfield  {author} {\bibinfo {author} {\bibfnamefont {L.}~\bibnamefont
  {Heisenberg}},\ }\href {https://doi.org/10.1016/j.physrep.2018.11.006}
  {\bibfield  {journal} {\bibinfo  {journal} {Phys. Rept.}\ }\textbf {\bibinfo
  {volume} {796}},\ \bibinfo {pages} {1} (\bibinfo {year} {2019})}\BibitemShut
  {NoStop}%
\bibitem [{\citenamefont {de~Azcarraga}\ \emph {et~al.}(2011)\citenamefont
  {de~Azcarraga}, \citenamefont {Kamimura},\ and\ \citenamefont
  {Lukierski}}]{azcarraga2011generalized}%
  \BibitemOpen
  \bibfield  {author} {\bibinfo {author} {\bibfnamefont {J.~A.}\ \bibnamefont
  {de~Azcarraga}}, \bibinfo {author} {\bibfnamefont {K.}~\bibnamefont
  {Kamimura}},\ and\ \bibinfo {author} {\bibfnamefont {J.}~\bibnamefont
  {Lukierski}},\ }\href {https://doi.org/10.1103/PhysRevD.83.124036} {\bibfield
   {journal} {\bibinfo  {journal} {Phys. Rev. D}\ }\textbf {\bibinfo {volume}
  {83}},\ \bibinfo {pages} {124036} (\bibinfo {year} {2011})}\BibitemShut
  {NoStop}%
\bibitem [{\citenamefont {Bacry}\ \emph
  {et~al.}(1970{\natexlab{a}})\citenamefont {Bacry}, \citenamefont {Combe},\
  and\ \citenamefont {Richard}}]{bacry1970group}%
  \BibitemOpen
  \bibfield  {author} {\bibinfo {author} {\bibfnamefont {H.}~\bibnamefont
  {Bacry}}, \bibinfo {author} {\bibfnamefont {P.}~\bibnamefont {Combe}},\ and\
  \bibinfo {author} {\bibfnamefont {J.~L.}\ \bibnamefont {Richard}},\ }\href
  {https://doi.org/10.1007/BF02725178} {\bibfield  {journal} {\bibinfo
  {journal} {Nuovo Cim. A}\ }\textbf {\bibinfo {volume} {67}},\ \bibinfo
  {pages} {267} (\bibinfo {year} {1970}{\natexlab{a}})}\BibitemShut {NoStop}%
\bibitem [{\citenamefont {Bacry}\ \emph
  {et~al.}(1970{\natexlab{b}})\citenamefont {Bacry}, \citenamefont {Combe},\
  and\ \citenamefont {Richard}}]{bacry1970group2}%
  \BibitemOpen
  \bibfield  {author} {\bibinfo {author} {\bibfnamefont {H.}~\bibnamefont
  {Bacry}}, \bibinfo {author} {\bibfnamefont {P.}~\bibnamefont {Combe}},\ and\
  \bibinfo {author} {\bibfnamefont {J.~L.}\ \bibnamefont {Richard}},\ }\href
  {https://doi.org/10.1007/BF02725375} {\bibfield  {journal} {\bibinfo
  {journal} {Nuovo Cim. A}\ }\textbf {\bibinfo {volume} {70}},\ \bibinfo
  {pages} {289} (\bibinfo {year} {1970}{\natexlab{b}})}\BibitemShut {NoStop}%
\bibitem [{\citenamefont {Schrader}(1972)}]{schrader1972maxwell}%
  \BibitemOpen
  \bibfield  {author} {\bibinfo {author} {\bibfnamefont {R.}~\bibnamefont
  {Schrader}},\ }\href {https://doi.org/10.1002/prop.19720201202} {\bibfield
  {journal} {\bibinfo  {journal} {Fortschr. Phys.}\ }\textbf {\bibinfo {volume}
  {20}},\ \bibinfo {pages} {701} (\bibinfo {year} {1972})}\BibitemShut
  {NoStop}%
\bibitem [{\citenamefont {Soroka}\ and\ \citenamefont
  {Soroka}(2005)}]{soroka2005tensor}%
  \BibitemOpen
  \bibfield  {author} {\bibinfo {author} {\bibfnamefont {D.~V.}\ \bibnamefont
  {Soroka}}\ and\ \bibinfo {author} {\bibfnamefont {V.~A.}\ \bibnamefont
  {Soroka}},\ }\href {https://doi.org/10.1016/j.physletb.2004.12.075}
  {\bibfield  {journal} {\bibinfo  {journal} {Phys. Lett. B}\ }\textbf
  {\bibinfo {volume} {607}},\ \bibinfo {pages} {302} (\bibinfo {year}
  {2005})}\BibitemShut {NoStop}%
\bibitem [{\citenamefont {Soroka}\ and\ \citenamefont
  {Soroka}(2012)}]{soroka2012gauge}%
  \BibitemOpen
  \bibfield  {author} {\bibinfo {author} {\bibfnamefont {D.~V.}\ \bibnamefont
  {Soroka}}\ and\ \bibinfo {author} {\bibfnamefont {V.~A.}\ \bibnamefont
  {Soroka}},\ }\href {https://doi.org/10.1016/j.physletb.2011.07.003}
  {\bibfield  {journal} {\bibinfo  {journal} {Phys. Lett. B}\ }\textbf
  {\bibinfo {volume} {707}},\ \bibinfo {pages} {160} (\bibinfo {year}
  {2012})}\BibitemShut {NoStop}%
\bibitem [{\citenamefont {Gomis}\ \emph {et~al.}(2009)\citenamefont {Gomis},
  \citenamefont {Kamimura},\ and\ \citenamefont
  {Lukierski}}]{Gomis:2009Deformations}%
  \BibitemOpen
  \bibfield  {author} {\bibinfo {author} {\bibfnamefont {J.}~\bibnamefont
  {Gomis}}, \bibinfo {author} {\bibfnamefont {K.}~\bibnamefont {Kamimura}},\
  and\ \bibinfo {author} {\bibfnamefont {J.}~\bibnamefont {Lukierski}},\ }\href
  {https://doi.org/10.1088/1126-6708/2009/08/039} {\bibfield  {journal}
  {\bibinfo  {journal} {JHEP}\ }\textbf {\bibinfo {volume} {08}},\ \bibinfo
  {pages} {039}}\BibitemShut {NoStop}%
\bibitem [{\citenamefont {Bonanos}\ and\ \citenamefont
  {Gomis}(2010)}]{bonanos2009infinite}%
  \BibitemOpen
  \bibfield  {author} {\bibinfo {author} {\bibfnamefont {S.}~\bibnamefont
  {Bonanos}}\ and\ \bibinfo {author} {\bibfnamefont {J.}~\bibnamefont
  {Gomis}},\ }\href {https://doi.org/10.1088/1751-8113/43/1/015201} {\bibfield
  {journal} {\bibinfo  {journal} {J. Phys. A}\ }\textbf {\bibinfo {volume}
  {43}},\ \bibinfo {pages} {015201} (\bibinfo {year} {2010})}\BibitemShut
  {NoStop}%
\bibitem [{\citenamefont {Bonanos}\ and\ \citenamefont
  {Gomis}(2009)}]{Bonanos:2009ChevalleyEilenberg}%
  \BibitemOpen
  \bibfield  {author} {\bibinfo {author} {\bibfnamefont {S.}~\bibnamefont
  {Bonanos}}\ and\ \bibinfo {author} {\bibfnamefont {J.}~\bibnamefont
  {Gomis}},\ }\href {https://doi.org/10.1088/1751-8113/42/14/145206} {\bibfield
   {journal} {\bibinfo  {journal} {J. Phys. A}\ }\textbf {\bibinfo {volume}
  {42}},\ \bibinfo {pages} {145206} (\bibinfo {year} {2009})}\BibitemShut
  {NoStop}%
\bibitem [{\citenamefont {Concha}\ and\ \citenamefont
  {Rodr\'\i{}guez}(2014)}]{concha2014maxwell}%
  \BibitemOpen
  \bibfield  {author} {\bibinfo {author} {\bibfnamefont {P.~K.}\ \bibnamefont
  {Concha}}\ and\ \bibinfo {author} {\bibfnamefont {E.~K.}\ \bibnamefont
  {Rodr\'\i{}guez}},\ }\href {https://doi.org/10.1016/j.nuclphysb.2014.07.022}
  {\bibfield  {journal} {\bibinfo  {journal} {Nucl. Phys. B}\ }\textbf
  {\bibinfo {volume} {886}},\ \bibinfo {pages} {1128} (\bibinfo {year}
  {2014})}\BibitemShut {NoStop}%
\bibitem [{\citenamefont {Cebecio\u{g}lu}\ and\ \citenamefont
  {Kibaro\u{g}lu}(2014)}]{cebeciouglu2014gauge}%
  \BibitemOpen
  \bibfield  {author} {\bibinfo {author} {\bibfnamefont {O.}~\bibnamefont
  {Cebecio\u{g}lu}}\ and\ \bibinfo {author} {\bibfnamefont {S.}~\bibnamefont
  {Kibaro\u{g}lu}},\ }\href {https://doi.org/10.1103/PhysRevD.90.084053}
  {\bibfield  {journal} {\bibinfo  {journal} {Phys. Rev. D}\ }\textbf {\bibinfo
  {volume} {90}},\ \bibinfo {pages} {084053} (\bibinfo {year}
  {2014})}\BibitemShut {NoStop}%
\bibitem [{\citenamefont {Cebecio\u{g}lu}\ and\ \citenamefont
  {Kibaro\u{g}lu}(2015)}]{cebeciouglu2015maxwell}%
  \BibitemOpen
  \bibfield  {author} {\bibinfo {author} {\bibfnamefont {O.}~\bibnamefont
  {Cebecio\u{g}lu}}\ and\ \bibinfo {author} {\bibfnamefont {S.}~\bibnamefont
  {Kibaro\u{g}lu}},\ }\href {https://doi.org/10.1016/j.physletb.2015.10.022}
  {\bibfield  {journal} {\bibinfo  {journal} {Phys. Lett. B}\ }\textbf
  {\bibinfo {volume} {751}},\ \bibinfo {pages} {131} (\bibinfo {year}
  {2015})}\BibitemShut {NoStop}%
\bibitem [{\citenamefont {Concha}\ \emph {et~al.}(2015)\citenamefont {Concha},
  \citenamefont {Rodr\'\i{}guez},\ and\ \citenamefont
  {Salgado}}]{concha2015generalized}%
  \BibitemOpen
  \bibfield  {author} {\bibinfo {author} {\bibfnamefont {P.~K.}\ \bibnamefont
  {Concha}}, \bibinfo {author} {\bibfnamefont {E.~K.}\ \bibnamefont
  {Rodr\'\i{}guez}},\ and\ \bibinfo {author} {\bibfnamefont {P.}~\bibnamefont
  {Salgado}},\ }\href {https://doi.org/10.1007/JHEP08(2015)009} {\bibfield
  {journal} {\bibinfo  {journal} {JHEP}\ }\textbf {\bibinfo {volume} {08}},\
  \bibinfo {pages} {009}}\BibitemShut {NoStop}%
\bibitem [{\citenamefont {Kibaro\u{g}lu}\ \emph {et~al.}(2019)\citenamefont
  {Kibaro\u{g}lu}, \citenamefont {\c{S}enay},\ and\ \citenamefont
  {Cebecio\u{g}lu}}]{kibarouglu2019maxwellSpecial}%
  \BibitemOpen
  \bibfield  {author} {\bibinfo {author} {\bibfnamefont {S.}~\bibnamefont
  {Kibaro\u{g}lu}}, \bibinfo {author} {\bibfnamefont {M.}~\bibnamefont
  {\c{S}enay}},\ and\ \bibinfo {author} {\bibfnamefont {O.}~\bibnamefont
  {Cebecio\u{g}lu}},\ }\href {https://doi.org/10.1142/S0217732319500160}
  {\bibfield  {journal} {\bibinfo  {journal} {Mod. Phys. Lett. A}\ }\textbf
  {\bibinfo {volume} {34}},\ \bibinfo {pages} {1950016} (\bibinfo {year}
  {2019})}\BibitemShut {NoStop}%
\bibitem [{\citenamefont {Kibaro\u{g}lu}\ and\ \citenamefont
  {Cebecio\u{g}lu}(2019)}]{kibarouglu2019super}%
  \BibitemOpen
  \bibfield  {author} {\bibinfo {author} {\bibfnamefont {S.}~\bibnamefont
  {Kibaro\u{g}lu}}\ and\ \bibinfo {author} {\bibfnamefont {O.}~\bibnamefont
  {Cebecio\u{g}lu}},\ }\href {https://doi.org/10.1140/epjc/s10052-019-7421-0}
  {\bibfield  {journal} {\bibinfo  {journal} {Eur. Phys. J. C}\ }\textbf
  {\bibinfo {volume} {79}},\ \bibinfo {pages} {898} (\bibinfo {year}
  {2019})}\BibitemShut {NoStop}%
\bibitem [{\citenamefont {Kibaro\u{g}lu}\ and\ \citenamefont
  {Cebecio\u{g}lu}(2020)}]{kibarouglu2020generalizedConformal}%
  \BibitemOpen
  \bibfield  {author} {\bibinfo {author} {\bibfnamefont {S.}~\bibnamefont
  {Kibaro\u{g}lu}}\ and\ \bibinfo {author} {\bibfnamefont {O.}~\bibnamefont
  {Cebecio\u{g}lu}},\ }\href {https://doi.org/10.1016/j.physletb.2020.135295}
  {\bibfield  {journal} {\bibinfo  {journal} {Phys. Lett. B}\ }\textbf
  {\bibinfo {volume} {803}},\ \bibinfo {pages} {135295} (\bibinfo {year}
  {2020})}\BibitemShut {NoStop}%
\bibitem [{\citenamefont {Kibaro\u{g}lu}\ and\ \citenamefont
  {Cebecio\u{g}lu}(2021)}]{kibarouglu2021gaugeAdS}%
  \BibitemOpen
  \bibfield  {author} {\bibinfo {author} {\bibfnamefont {S.}~\bibnamefont
  {Kibaro\u{g}lu}}\ and\ \bibinfo {author} {\bibfnamefont {O.}~\bibnamefont
  {Cebecio\u{g}lu}},\ }\href {https://doi.org/10.1142/S0218271821500759}
  {\bibfield  {journal} {\bibinfo  {journal} {Int. J. Mod. Phys. D}\ }\textbf
  {\bibinfo {volume} {30}},\ \bibinfo {pages} {2150075} (\bibinfo {year}
  {2021})}\BibitemShut {NoStop}%
\bibitem [{\citenamefont {Cebecio\u{g}lu}\ and\ \citenamefont
  {Kibaro\u{g}lu}(2021)}]{cebeciouglu2021maxwellMetricAffine}%
  \BibitemOpen
  \bibfield  {author} {\bibinfo {author} {\bibfnamefont {O.}~\bibnamefont
  {Cebecio\u{g}lu}}\ and\ \bibinfo {author} {\bibfnamefont {S.}~\bibnamefont
  {Kibaro\u{g}lu}},\ }\href {https://doi.org/10.1140/epjc/s10052-021-09685-6}
  {\bibfield  {journal} {\bibinfo  {journal} {Eur. Phys. J. C}\ }\textbf
  {\bibinfo {volume} {81}},\ \bibinfo {pages} {900} (\bibinfo {year}
  {2021})}\BibitemShut {NoStop}%
\bibitem [{\citenamefont {Baykal}\ and\ \citenamefont
  {Delice}(2011)}]{Baykal:2010Unified}%
  \BibitemOpen
  \bibfield  {author} {\bibinfo {author} {\bibfnamefont {A.}~\bibnamefont
  {Baykal}}\ and\ \bibinfo {author} {\bibfnamefont {O.}~\bibnamefont
  {Delice}},\ }\href {https://doi.org/10.1088/0264-9381/28/1/015014} {\bibfield
   {journal} {\bibinfo  {journal} {Class. Quant. Grav.}\ }\textbf {\bibinfo
  {volume} {28}},\ \bibinfo {pages} {015014} (\bibinfo {year}
  {2011})}\BibitemShut {NoStop}%
\bibitem [{\citenamefont {Baykal}\ and\ \citenamefont
  {Delice}(2013)}]{Baykal:2013MultiScalar}%
  \BibitemOpen
  \bibfield  {author} {\bibinfo {author} {\bibfnamefont {A.}~\bibnamefont
  {Baykal}}\ and\ \bibinfo {author} {\bibfnamefont {O.}~\bibnamefont
  {Delice}},\ }\href {https://doi.org/10.1103/PhysRevD.88.084041} {\bibfield
  {journal} {\bibinfo  {journal} {Phys. Rev. D}\ }\textbf {\bibinfo {volume}
  {88}},\ \bibinfo {pages} {084041} (\bibinfo {year} {2013})}\BibitemShut
  {NoStop}%
\bibitem [{\citenamefont {Cardenas}\ \emph {et~al.}(2022)\citenamefont
  {Cardenas}, \citenamefont {Diaz}, \citenamefont {Salgado},\ and\
  \citenamefont {Salgado}}]{Cardenas:2022GeneralizedEinstein}%
  \BibitemOpen
  \bibfield  {author} {\bibinfo {author} {\bibfnamefont {L.}~\bibnamefont
  {Cardenas}}, \bibinfo {author} {\bibfnamefont {J.}~\bibnamefont {Diaz}},
  \bibinfo {author} {\bibfnamefont {P.}~\bibnamefont {Salgado}},\ and\ \bibinfo
  {author} {\bibfnamefont {D.}~\bibnamefont {Salgado}},\ }\href
  {https://doi.org/10.1016/j.nuclphysb.2022.115943} {\bibfield  {journal}
  {\bibinfo  {journal} {Nucl. Phys. B}\ }\textbf {\bibinfo {volume} {984}},\
  \bibinfo {pages} {115943} (\bibinfo {year} {2022})}\BibitemShut {NoStop}%
\bibitem [{\citenamefont {Pe\~nafiel}\ and\ \citenamefont
  {Ravera}(2018)}]{Penafiel:2018AdS}%
  \BibitemOpen
  \bibfield  {author} {\bibinfo {author} {\bibfnamefont {D.~M.}\ \bibnamefont
  {Pe\~nafiel}}\ and\ \bibinfo {author} {\bibfnamefont {L.}~\bibnamefont
  {Ravera}},\ }\href {https://doi.org/10.1140/epjc/s10052-018-6421-9}
  {\bibfield  {journal} {\bibinfo  {journal} {Eur. Phys. J. C}\ }\textbf
  {\bibinfo {volume} {78}},\ \bibinfo {pages} {945} (\bibinfo {year}
  {2018})}\BibitemShut {NoStop}%
\bibitem [{\citenamefont {Durka}\ \emph {et~al.}(2011)\citenamefont {Durka},
  \citenamefont {Kowalski-Glikman},\ and\ \citenamefont
  {Szczachor}}]{durka2011gauged}%
  \BibitemOpen
  \bibfield  {author} {\bibinfo {author} {\bibfnamefont {R.}~\bibnamefont
  {Durka}}, \bibinfo {author} {\bibfnamefont {J.}~\bibnamefont
  {Kowalski-Glikman}},\ and\ \bibinfo {author} {\bibfnamefont {M.}~\bibnamefont
  {Szczachor}},\ }\href {https://doi.org/10.1142/S0217732311037078} {\bibfield
  {journal} {\bibinfo  {journal} {Mod. Phys. Lett. A}\ }\textbf {\bibinfo
  {volume} {26}},\ \bibinfo {pages} {2689} (\bibinfo {year}
  {2011})}\BibitemShut {NoStop}%
\bibitem [{\citenamefont {Izaurieta}\ \emph {et~al.}(2006)\citenamefont
  {Izaurieta}, \citenamefont {Rodriguez},\ and\ \citenamefont
  {Salgado}}]{Izaurieta:2006ExpandingLie}%
  \BibitemOpen
  \bibfield  {author} {\bibinfo {author} {\bibfnamefont {F.}~\bibnamefont
  {Izaurieta}}, \bibinfo {author} {\bibfnamefont {E.}~\bibnamefont
  {Rodriguez}},\ and\ \bibinfo {author} {\bibfnamefont {P.}~\bibnamefont
  {Salgado}},\ }\href {https://doi.org/10.1063/1.2390659} {\bibfield  {journal}
  {\bibinfo  {journal} {J. Math. Phys.}\ }\textbf {\bibinfo {volume} {47}},\
  \bibinfo {pages} {123512} (\bibinfo {year} {2006})}\BibitemShut {NoStop}%
\bibitem [{\citenamefont {Salgado}\ and\ \citenamefont
  {Salgado}(2014)}]{Salgado:2014soD1}%
  \BibitemOpen
  \bibfield  {author} {\bibinfo {author} {\bibfnamefont {P.}~\bibnamefont
  {Salgado}}\ and\ \bibinfo {author} {\bibfnamefont {S.}~\bibnamefont
  {Salgado}},\ }\href {https://doi.org/10.1016/j.physletb.2013.11.009}
  {\bibfield  {journal} {\bibinfo  {journal} {Phys. Lett. B}\ }\textbf
  {\bibinfo {volume} {728}},\ \bibinfo {pages} {5} (\bibinfo {year}
  {2014})}\BibitemShut {NoStop}%
\bibitem [{\citenamefont {Capozziello}\ \emph {et~al.}(2007)\citenamefont
  {Capozziello}, \citenamefont {Cianci}, \citenamefont {Stornaiolo},\ and\
  \citenamefont {Vignolo}}]{Capozziello:2007f(R)gravity}%
  \BibitemOpen
  \bibfield  {author} {\bibinfo {author} {\bibfnamefont {S.}~\bibnamefont
  {Capozziello}}, \bibinfo {author} {\bibfnamefont {R.}~\bibnamefont {Cianci}},
  \bibinfo {author} {\bibfnamefont {C.}~\bibnamefont {Stornaiolo}},\ and\
  \bibinfo {author} {\bibfnamefont {S.}~\bibnamefont {Vignolo}},\ }\href
  {https://doi.org/10.1088/0264-9381/24/24/015} {\bibfield  {journal} {\bibinfo
   {journal} {Class. Quant. Grav.}\ }\textbf {\bibinfo {volume} {24}},\
  \bibinfo {pages} {6417} (\bibinfo {year} {2007})}\BibitemShut {NoStop}%
\bibitem [{\citenamefont {Capozziello}\ \emph {et~al.}(2008)\citenamefont
  {Capozziello}, \citenamefont {Cianci}, \citenamefont {Stornaiolo},\ and\
  \citenamefont {Vignolo}}]{Capozziello:2008f(R)gravity}%
  \BibitemOpen
  \bibfield  {author} {\bibinfo {author} {\bibfnamefont {S.}~\bibnamefont
  {Capozziello}}, \bibinfo {author} {\bibfnamefont {R.}~\bibnamefont {Cianci}},
  \bibinfo {author} {\bibfnamefont {C.}~\bibnamefont {Stornaiolo}},\ and\
  \bibinfo {author} {\bibfnamefont {S.}~\bibnamefont {Vignolo}},\ }\href
  {https://doi.org/10.1142/S0219887808003053} {\bibfield  {journal} {\bibinfo
  {journal} {Int. J. Geom. Meth. Mod. Phys.}\ }\textbf {\bibinfo {volume}
  {5}},\ \bibinfo {pages} {765} (\bibinfo {year} {2008})}\BibitemShut {NoStop}%
\bibitem [{\citenamefont {de~Azc{\'a}rraga}\ \emph {et~al.}(2013)\citenamefont
  {de~Azc{\'a}rraga}, \citenamefont {Kamimura},\ and\ \citenamefont
  {Lukierski}}]{Azcarraga2013maxwellApplication}%
  \BibitemOpen
  \bibfield  {author} {\bibinfo {author} {\bibfnamefont {J.~A.}\ \bibnamefont
  {de~Azc{\'a}rraga}}, \bibinfo {author} {\bibfnamefont {K.}~\bibnamefont
  {Kamimura}},\ and\ \bibinfo {author} {\bibfnamefont {J.}~\bibnamefont
  {Lukierski}},\ }in\ \href {https://doi.org/10.1142/S2010194513011604} {\emph
  {\bibinfo {booktitle} {Int. J. Mod. Phys. Conf. Ser.}}},\ Vol.~\bibinfo
  {volume} {23}\ (\bibinfo {year} {2013})\ pp.\ \bibinfo {pages}
  {350--356}\BibitemShut {NoStop}%
\end{thebibliography}%

\end{document}